\documentclass[useAMS,usenatbib]{mn2e}
\usepackage{times}
\usepackage[fleqn]{amsmath}
\usepackage{amssymb}
\usepackage{graphicx}
\usepackage{subfigure}
\usepackage{float}
\usepackage{relsize}
\usepackage{color}
\usepackage{gensymb}
\usepackage[mathscr]{euscript}

\title[Dust migration and toroidal vortices]{Toroidal vortices as a solution to the dust migration problem}\author[Pablo Lor\'en-Aguilar and Matthew R. Bate]{Pablo Lor\'en-Aguilar$^{1}$\thanks{E-mail:
pablo@astro.ex.ac.uk},{Matthew R. Bate$^{1}$\thanks{E-mail: mbate@astro.ex.ac.uk}} \\ $^{1}$ School of Physics and Astronomy, University of Exeter, Stocker Road, Exeter EX4 4QL, United Kingdom}

\date{Accepted 2015 December 17; Received 2015 December 17; in original form 2015 December 08}

\begin{document}

\pagerange{\pageref{firstpage}--\pageref{lastpage}} \pubyear{???} \maketitle
\label{firstpage}

\begin{abstract}
In an earlier letter, we reported that dust settling in protoplanetary discs may lead to a dynamical dust-gas instability that produces global toroidal vortices.  In this letter, we investigate the evolution of a dusty protoplanetary disc with two different dust species (1 mm and 50 cm dust grains), under the presence of the instability. We show how toroidal vortices, triggered by the interaction of mm grains with the gas, stop the radial migration of metre-sized dust, potentially offering a natural and efficient solution to the dust migration problem.
\end{abstract}


\begin{figure*} \centering
\includegraphics[width=88mm]{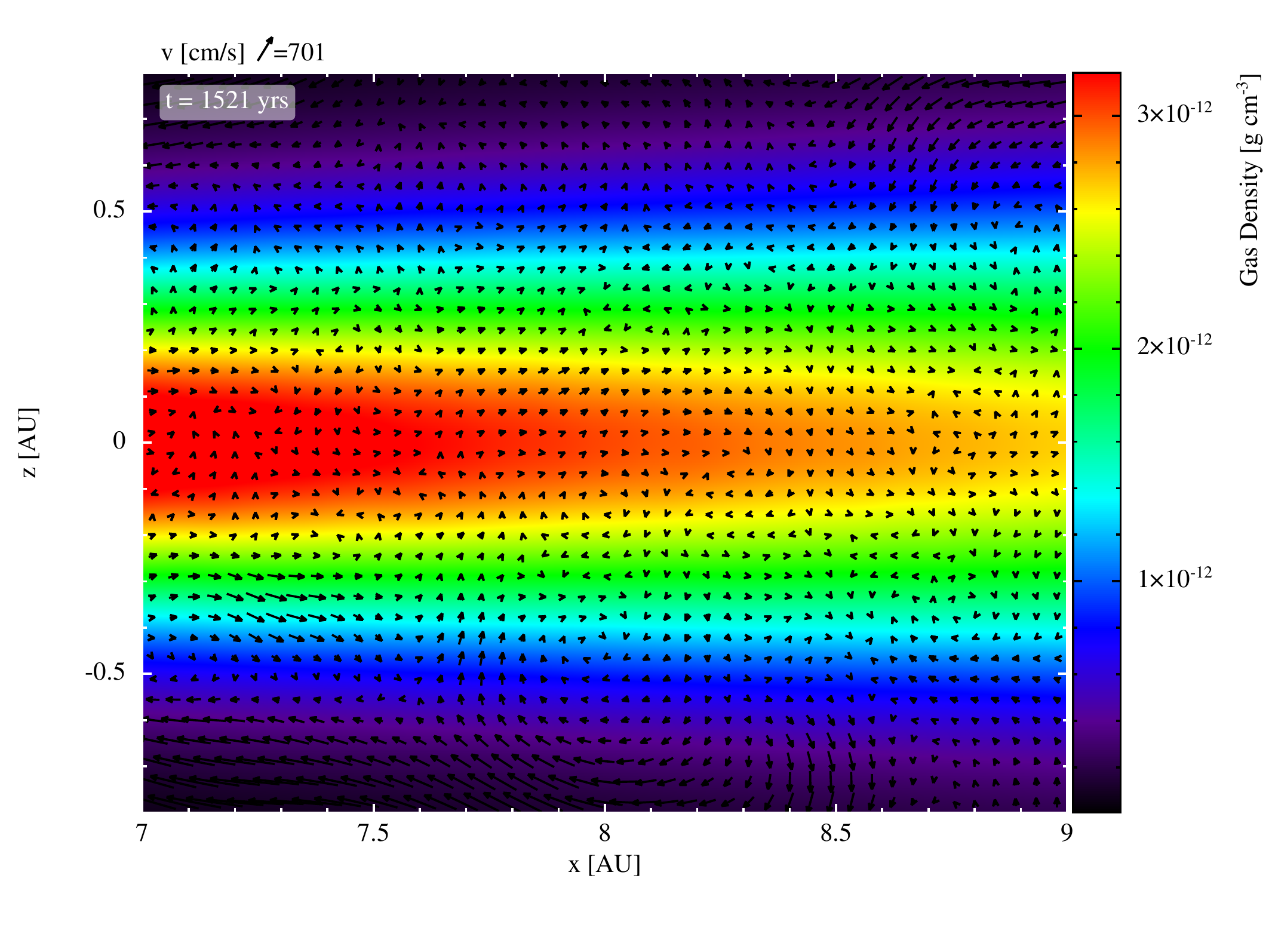}
\includegraphics[width=88mm]{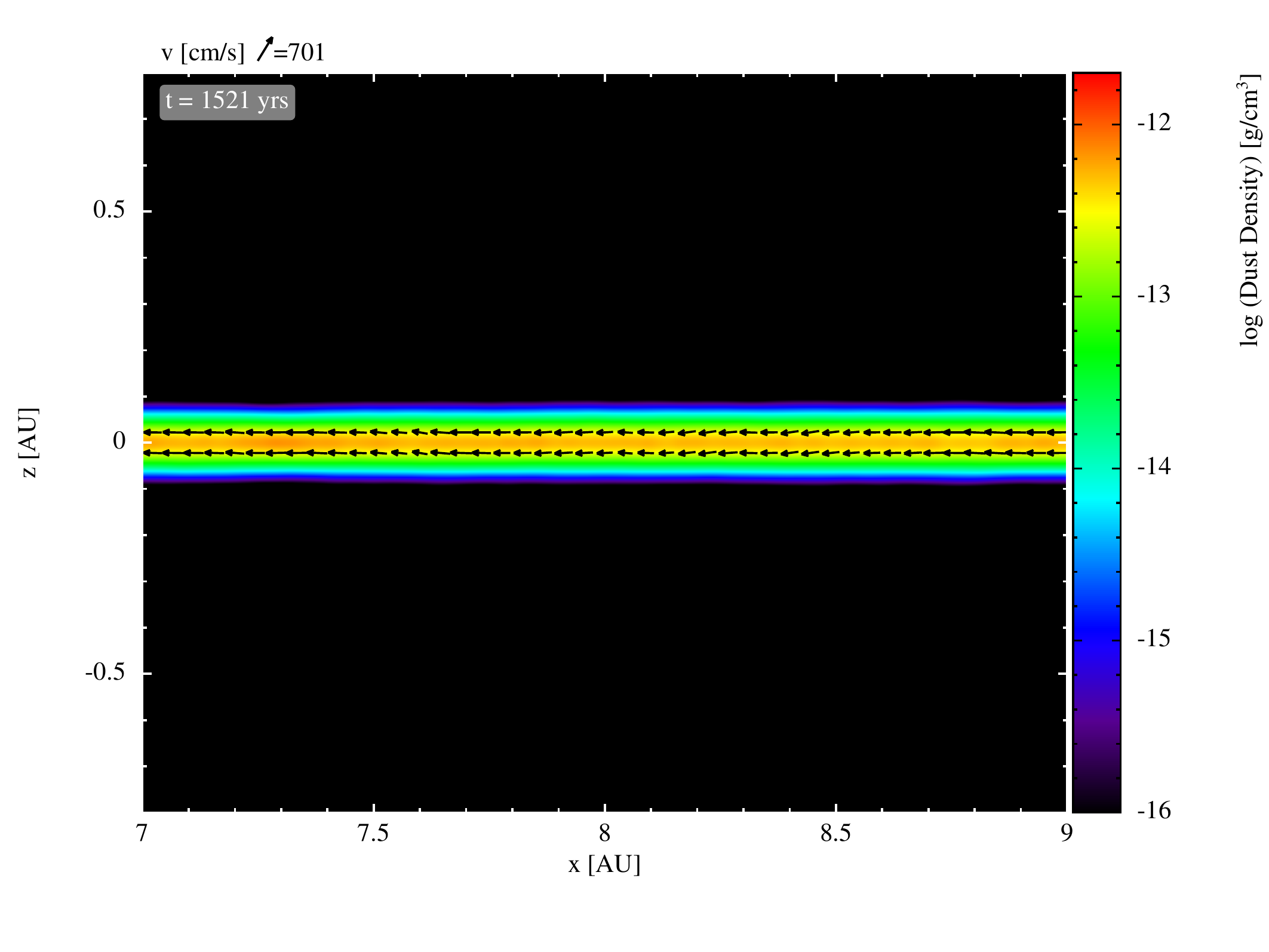} \vspace{-1cm}\\
\includegraphics[width=88mm]{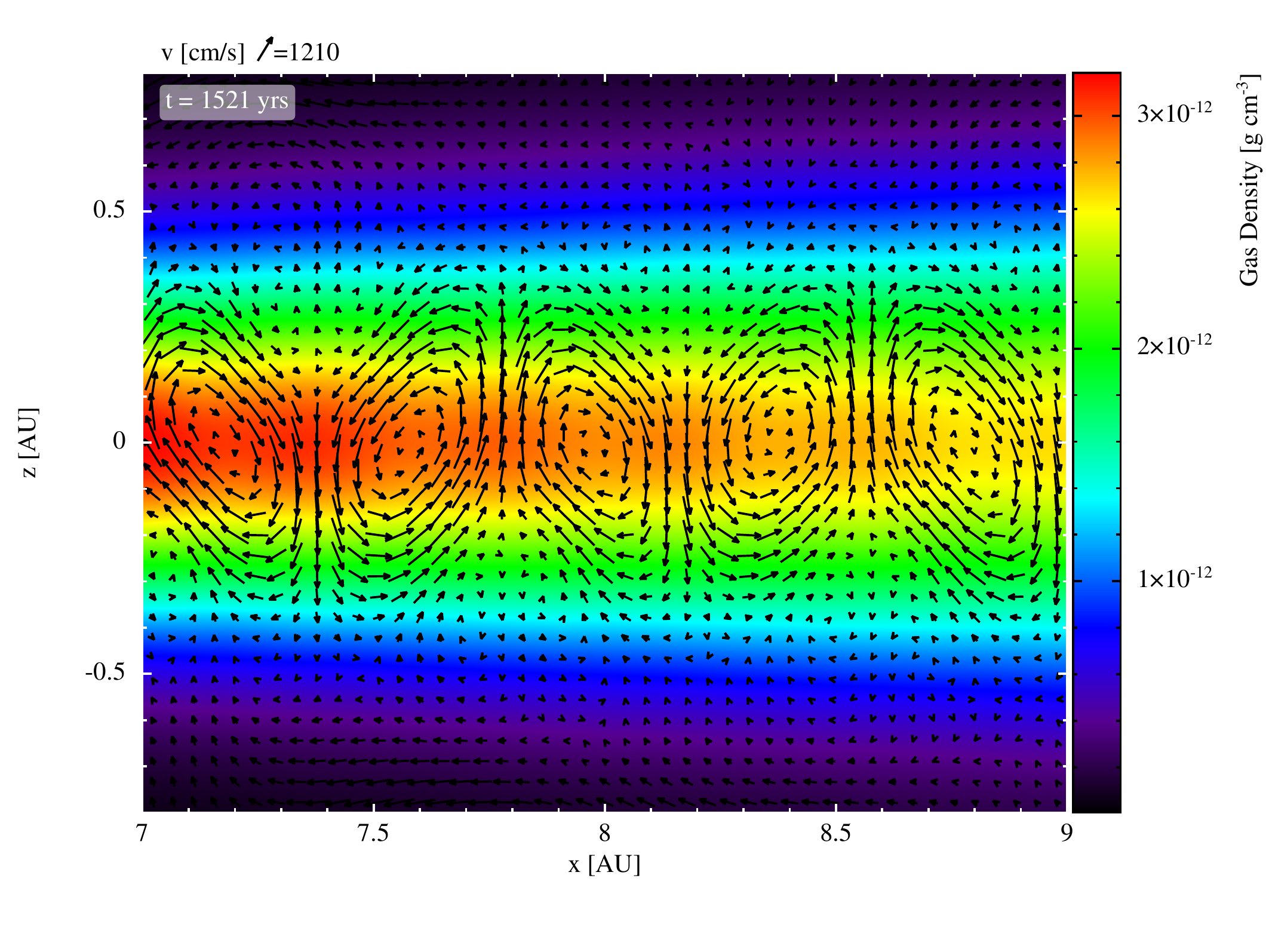}
\includegraphics[width=88mm]{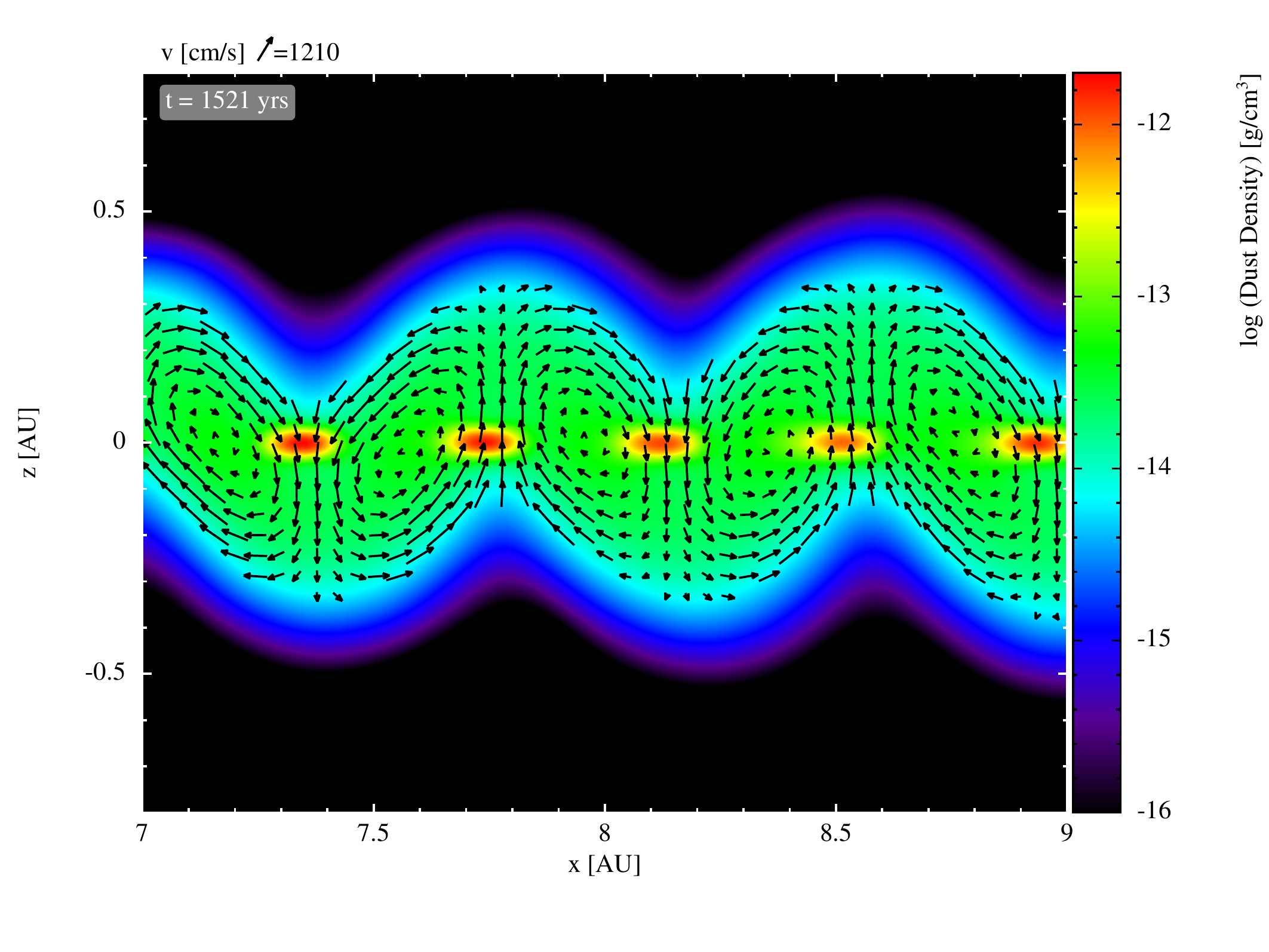} \vspace{-0.75cm}
\caption{Top panels:  The azimuthally-averaged gas (top left) and dust (top right) densities with velocity vectors are shown for the calculation with only 50-cm dust. The dust quickly settles to the mid-plane, and then rapidly migrates inward due to the drag interaction with the gas.  Bottom panels: The azimuthally-averaged gas (bottom left) and dust (bottom right) densities with velocity vectors for the calculation including both 1-mm and 50-cm dust, after the onset of the instability. The 1-mm dust grains are entrained in the gas vortices, forming a characteristic wave-like structure in the disc cross-section. On the contrary, the 50-cm dust grains become trapped in the inter-vortex regions, stopping their inward migration. They correspond to the high-density dust concentrations at the disc mid-plane. Note that the gas density is shown in linear scale so that the radial pressure variations at the mid-plane can be seen in the bottom left panel, while the dust density is shown in logarithmic scale.  Also, the velocity vectors are scaled differently for the two calculations.}
\label{Fig1}
\end{figure*}

\begin{figure*}
\includegraphics[width=85mm]{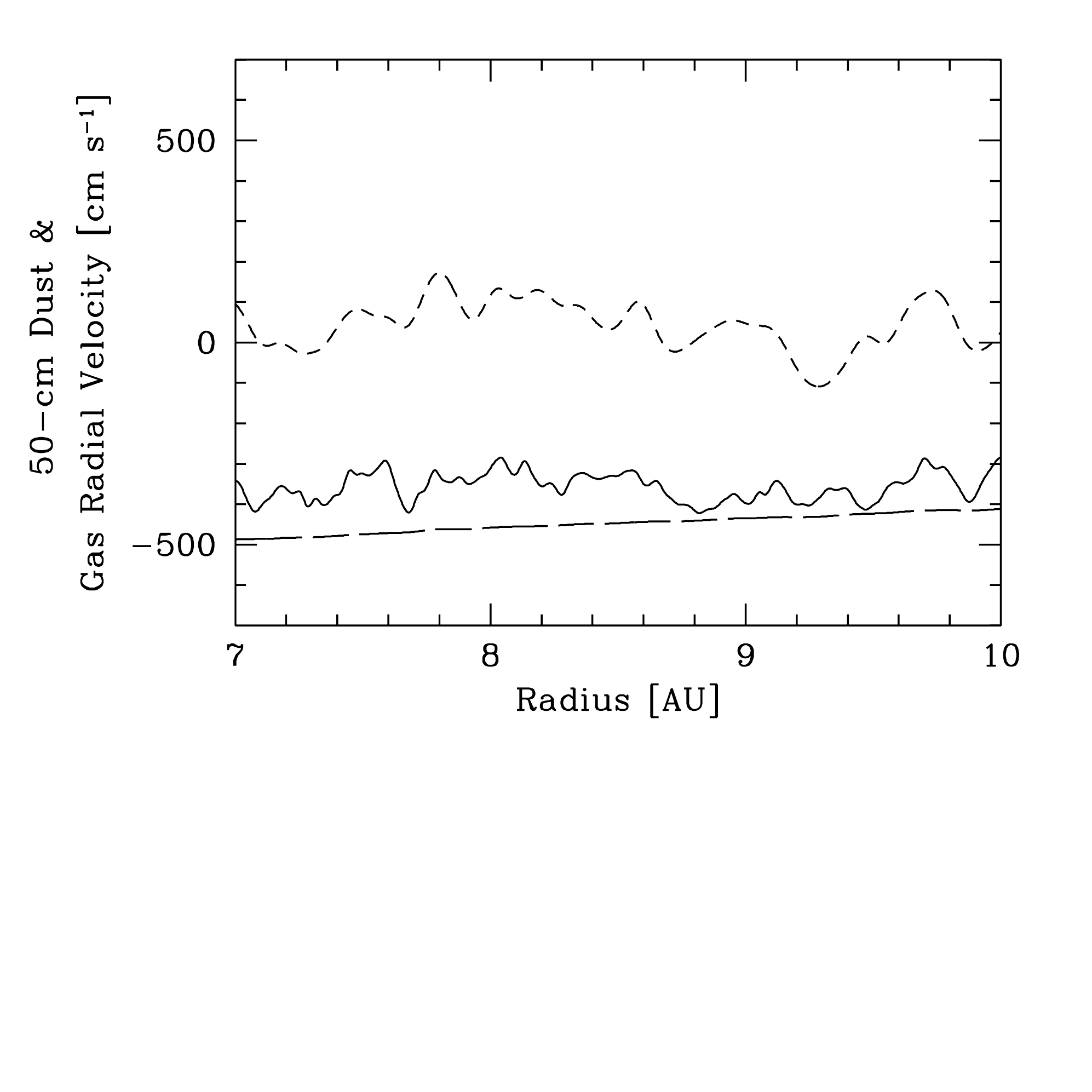} \vspace{-3cm}\\
\includegraphics[width=85mm]{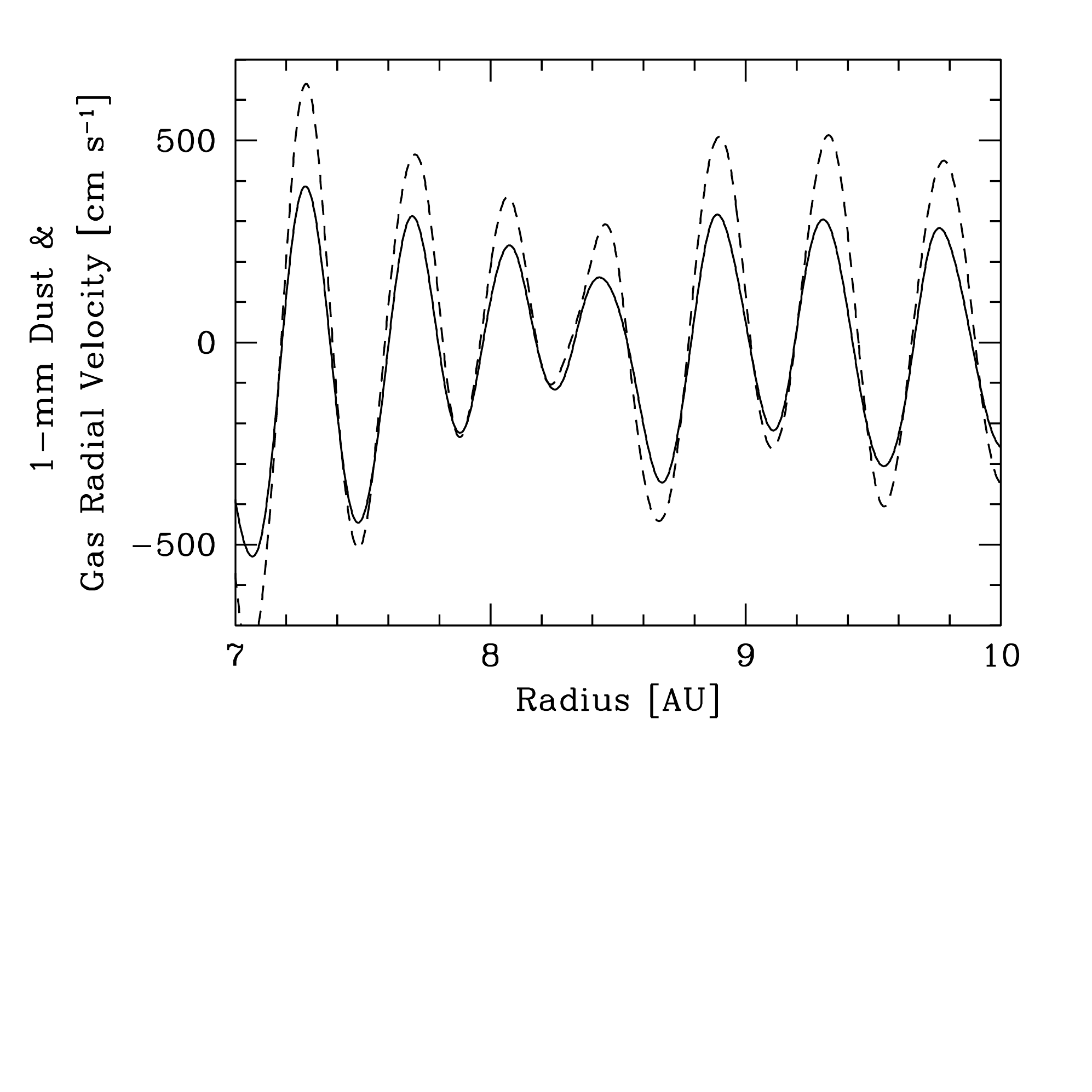}
\includegraphics[width=85mm]{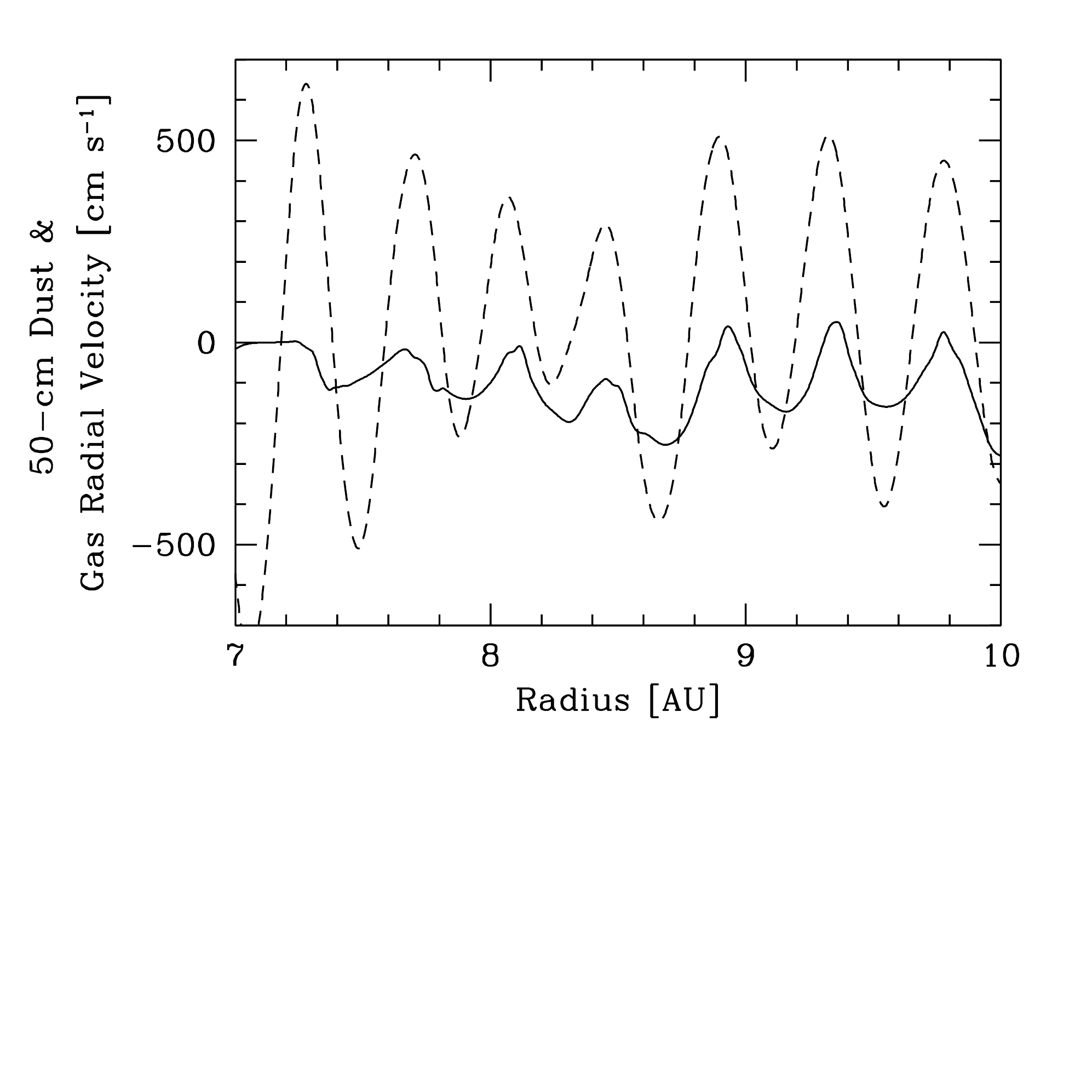} \vspace{-3cm}
\caption{Azimuthally-averaged gas (short-dashed lines) and dust (solid lines) radial velocity profiles at the mid-plane for the 50-cm case (top panel), and 1-mm + 50-cm case after the onset of the instability (bottom panels). In the 50-cm case (top panel), the dust migrates inwards at a rate close to that expected from equation \ref{vr} (long-dashed line). However, when both 1-mm and 50-cm grains are included, the radial velocity of the 50-cm dust grains is drastically reduced near gas density maxima (bottom right). On the contrary, 1-mm dust grains (bottom left) remain strongly coupled to the gas. Note that the Keplerian velocity at 8.5~AU is $1.0\times 10^6$~cm~s$^{-1}$.}
\label{Fig2}
\end{figure*}

\begin{figure*}
\includegraphics[width=85mm]{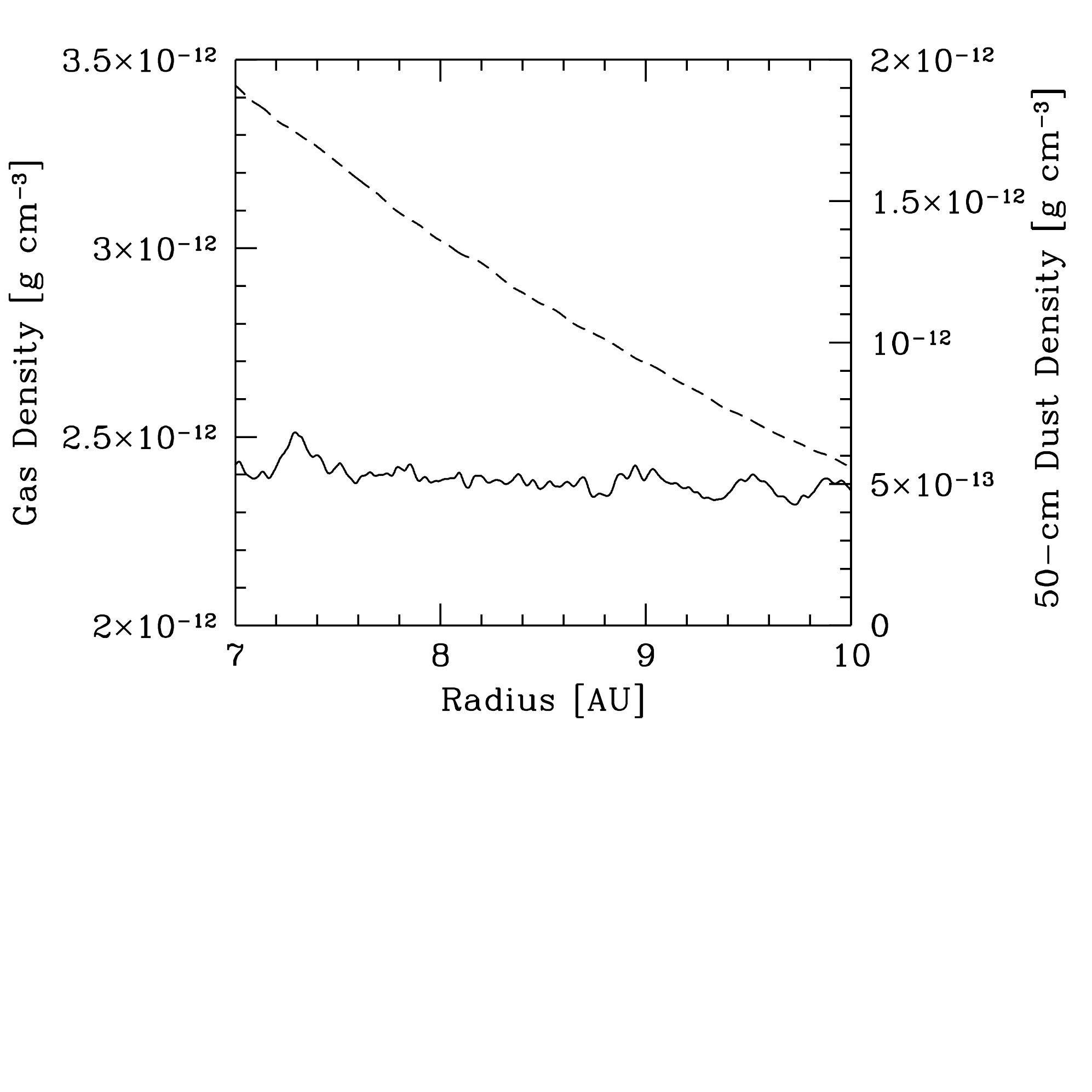} \vspace{-3cm} \\
\includegraphics[width=85mm]{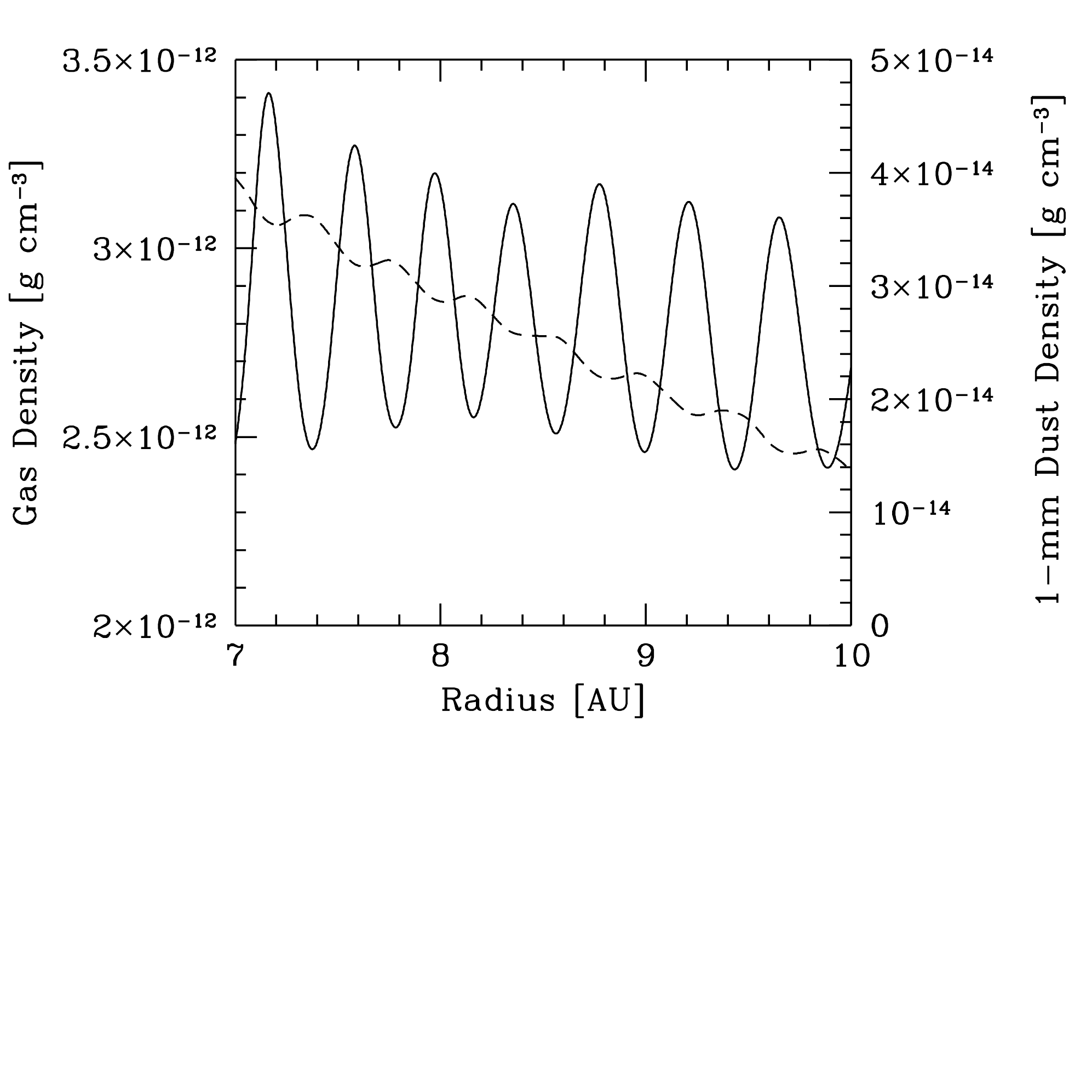}
\includegraphics[width=85mm]{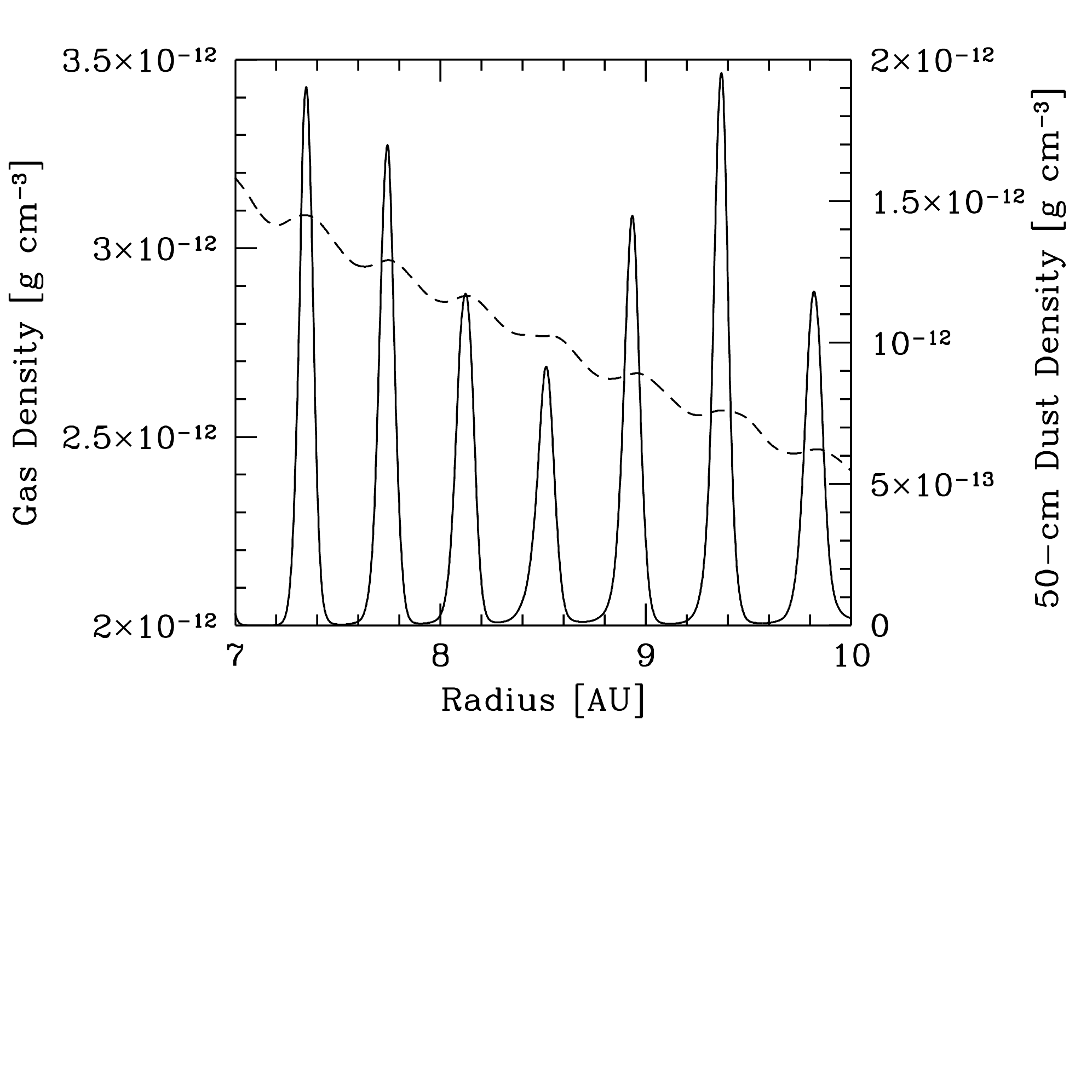} \vspace{-3cm}
\caption{Azimuthally-averaged gas (short-dashed lines) and dust (solid lines) radial density profiles at the mid-plane for the 50-cm case (upper panel), and the 1-mm + 50-cm case after the onset of the instability (bottom panels).  In the 50-cm case, the gas radial density profile remains smooth and the dust density shows no significant structure.  However, when both 1-mm and 50-cm grains are included, the 50-cm dust grains pile up at gas density maxima (bottom right), while the 1-mm dust grains (bottom left) concentrate at the centre of the vortices, where the gas density has local minima. }
\label{Fig3}
\end{figure*}

\section{Introduction}

According to the presently accepted paradigm, planets form through some kind of aggregation process of the dust grains present in protoplanetary discs. Such aggregation is proposed to lead to the formation of small planetary embryos of order a km in size, so-called planetesimals. However, this scenario suffers some severe problems that challenge our current theoretical understanding \citep[e.g.][]{CY10}. Dust migration is undoubtedly one of the most severe obstacles for the present planet formation paradigm. In laminar protoplanetary discs, dust grains bigger than a few centimetres in size suffer such a strong inward migration due to aero-dynamic drag that, unless some stopping mechanism is found, they would fall into the central star in less than a few thousand years \citep{Wei77}. Several mechanisms have been proposed in order to prevent such a strong migration, the vast majority of them involving some sort of dust trapping mechanism. 

One possible way to trap solids in a disc that has been thoroughly investigated during the last years is the streaming instability \citep{YG05, YJ07, JBL11, YJ14}. Triggered by local enhancements of the dust-to-gas ratio and a strong dust-gas drag, it can create dust clumps that may be the progenitors of planetesimals, although it is not completely clear yet how this mechanism will lead to the formation of planets in a realistic scenario \citep[see, for example, ][]{DD14}. Another recently proposed mechanism is the formation of dusty rings in protoplanetary discs due to the photoelectric instability \citep{LK13}. Unfortunately this instability can only operate in low mass, optically thin discs.  It is consequently difficult to use as a universal mechanism to prevent dust migration. Vortices have also been proposed as potentially effective mechanisms to trap dust \citep{BargeSommeria1995,TBD96,KB06}. In recent years, two different vortex-production mechanisms have been intensively investigated: the baroclinic instability \citep{GodonLivio1999,KB03,JAB04,Klahr04,PJS07,RKL15} capable of producing vortices due to the presence of an hydrodynamical instability driven by an entropy gradient in the disc, and the Rossby wave instability \citep{Lea99,Lea01,VT06,IB06,Lea09,Mehea12} driven by the presence of a pressure extremum. Recently, a new vortex-producing scenario was proposed by \cite{LB15a}. A new type of instability in protoplanetary discs, caused by the settling of intermediately-sized dust grains ($\sim$ 1 mm), was shown to be capable of developing global toroidal vortices in the gas component of the disc. Due to the strong coupling between the dust and gas components, dust grains become trapped in the vortices forming a set of dust rings throughout the disc. After saturation of the instability, the radial wavelength of the density perturbations is roughly equal to the thickness of the dust layer (see \cite{LB15a} for more details). 

The resulting perturbed disc configuration offers a promising solution to the dust migration problem.  Due to the circulation induced by the vortices, high density (and hence pressure) regions are created, and they may potentially act as a dust trap. Interestingly enough, a model that strongly resembles this configuration was thoroughly studied by \cite{Pea12}. According to their work, a sinusoidal perturbation in the radial gas surface density profile of a protoplanetary disc would allow dust grains at large radii to stop their inward migration and grow in size, thereby explaining mm-observations of young discs in the Taurus, Ophiuchus and Orion Nebula Cluster star-forming regions. The authors discussed various possible explanations for the origin of such density perturbations, such as the Rossby wave instability or zonal flows due to the magnetorotational instability. 

In this letter, we show that the instability reported by \cite{LB15a} naturally produces such radial perturbations of the gas surface density, and thereby offers a potential solution to the dust migration problem. We report the results from three-dimensional numerical simulations of dust settling in a protoplanetary disc using a two-fold population of dust grains, with sizes of 1 mm and 50 cm, along with the results from a control calculation that only contains 50-cm grains. The paper is organised as follows. In Section \ref{numerical}, we describe the numerical method and initial conditions that we use for hydrodynamical simulations of dusty protoplanetary discs.  Results from, and discussion of, our calculations are presented in Section \ref{results}, and in Section \ref{conclusions} we give our conclusions.

\section{Numerical method}
\label{numerical}

To perform numerical simulations of dusty protoplanetary discs, we have used the three-dimensional smoothed particle hydrodynamics (SPH) \citep{Lucy1977, GM77} code, {\tt sphNG} \citep{Bate95,Aea12}, which has been used to study a variety of astrophysical fluid dynamical systems. For the simulations performed for this letter, the code uses two populations of SPH particles: gas and dust.  The latter only evolves under the action of gravity and a drag force.  In order to perform the time-integration of the dust-gas drag force, we use the semi-implicit integration method of \cite{LB14}, including the improvement of \cite{LB15b}. 

The simulations follow the time evolution of locally-isothermal circumstellar gas/dust discs. The central star was modelled as a gravitational point mass, $M_{*} = 1~{\rm M}_{\odot}$, the mass of the gas component between the inner and outer radii was set to be $\approx 0.003 ~{\rm M}_\odot$ (similar to the minimum-mass Solar nebula), and the mass of the dust component was determined by the initial dust-to-gas ratio $\epsilon = 0.01$. The gaseous component of the disc had an outer radius of $r = 16.1$~AU, with an inner boundary at $r = 2.6$~AU, the latter chosen to avoid having particles with excessively short integration time-steps. The radial boundaries of the gas were treated using ghost particles.  The modelled discs were locally-isothermal with radial temperature profiles $T\propto r^{-1}$, vertical gas scale heights $H/r=0.04$, and uniform in surface density.  Self-gravity was neglected.  The calculations were performed with a resolution of $N_{\rm p} = 3\times 10^6$ SPH particles. In each case, approximately 1/3 of the particles were dust particles.  A purely gaseous disc was set up and relaxed, before the dust was added.  The initial dust-to-gas ratios were small enough to avoid significant perturbations of the gas hydrostatic equilibrium when the dust was added. The dust grains were introduced into the discs following the relaxed gas density distribution, but with minimum and maximum orbital radii of $r = 5.2$ AU and $r = 15.6$ AU, respectively.  The dust grains were assumed to be spherical, with intrinsic dust grain densities of $\hat{\rho}_{\rm D}=3.0$ g~cm$^{-3}$.  We performed two simulations: one with only dust grains 50~cm in radius, and the other with mixture of dust grains with radii of 1 mm and 50 cm. In both calculations, $\epsilon = 0.01$, but in the latter each type of dust comprised half of the total dust mass and was modelled using half of the total number of SPH dust particles.

\section{Results and discussion}
\label{results}

In Fig. \ref{Fig1}, we present results from the two simulations. In the upper panels, we show gas (top left) and dust (top right) density renderings of the simulation with only one dust population (50 cm). In the lower panels, the gas (bottom left) and dust (bottom right) density rendering of the protoplanetary disc simulation with two different dust grain populations (1-mm and 50-cm dust grains) is displayed. As expected, if only the 50-cm dust grains are present the dust-gas instability doesn't develop, and the gas component remains stable (top left). Vortices do not develop because the coupling between the dust and gas is weak, the dust settles quickly, and the instability does not have sufficient time to develop. After settling, the 50-cm dust migrates inward as expected due to drag with the gas, which has sub-Keplerian rotation.  If the gas radial pressure profile at the mid-plane can be expressed as $P \propto r^{-n}$, the gas will rotate at a slightly sub-Keplerian velocity given by
\begin{equation}
{\it v}_{\rm \phi, gas} = {\it v}_{\rm K}(1-\eta)^{1/2},
\end{equation}
where $\eta = nc_{\rm s}^2/v_{\rm K}^2$, $c_{\rm s}$ is the gas sound speed and ${\it v}_{\rm K}$ is the Keplerian velocity. Then, the friction between the gas and the dust (which rotates at the Keplerian velocity in the absence of drag) induces a radial drift of dust particles with velocity \citep[see, for example, ][]{Arm10}
\begin{equation}
{\it v}_{\rm r} = \frac{{\rm St}^{-1}{\it v}_{\rm r, gas} - \eta{\it v}_{\rm K}}{{\rm St} + {\rm St}^{-1}}, \label{vr}
\end{equation}
where ${\rm St} = \Omega_{\rm K}t_{\rm s}$ is the Stokes number of the dust grains, $\Omega_{\rm K}$ is the Keplerian angular velocity, $t_{\rm s} = \hat{\rho}_{\rm D}s/{\it v}_{\rm th}\rho_{\rm G}$ is the dust stopping time in the Epstein regime, $\hat{\rho}_{\rm D}$ is the dust grain intrinsic density, $\rho_{\rm G}$ is the gas density, ${\it s}$ is the dust grain radius, ${\it v}_{\rm th}= c_{\rm s}\sqrt{8/\pi}$, and ${\it v}_{\rm r,gas}$ is the radial velocity of the gas. In the top panel of Fig. \ref{Fig2} the radial velocity of the dust grains as a function of the local Keplerian velocity is plotted (solid line) versus the analytical expectation obtained by using equation \ref{vr} (long-dashed line) with our simulation parameters (${\it s} = 50~{\rm cm}$, ${\it n}=2$) and ${\it v}_{\rm r,gas}$ and $\rho_{\rm G}$ taken from the simulation. As can be seen, there is reasonable agreement between the simulated radial migration speed of the dust and the prediction of equation \ref{vr} --- on average the migration rate is about 15\% slower than the analytical prediction.  There is some numerical velocity dispersion, but note that the Keplerian velocity at these radii is $v_{\rm K} \approx 1\times 10^6$~cm~s$^{-1}$, and the sound speed is $c_{\rm s} \approx 4\times 10^4$~cm~s$^{-1}$ (i.e. much larger than the numerical dispersion).  This demonstrates that our method produces the expected radial migration of large dust grains in a laminar disc. In the top panel of Fig. \ref{Fig3} the dust and gas azimuthally-averaged radial density profiles are shown at the mid-plane. The gas density profile is smooth, as expected for a disc in hydrostatic balance, whereas the dust density profile shows small fluctuations with no systematic structure.

If 1-mm dust grains are included in the simulation, global coherent vortices appear (bottom left panel of Fig.~\ref{Fig1}) in the gas component. Due to the strong drag suffered from the gas (${\rm St} ~\approx 1\times 10^{-2}$) the 1-mm grains accumulate in the vortices, forming the characteristic wave-like structure that can be appreciated in the bottom right panel of Fig.~\ref{Fig1}. As in \cite{LB15a}, the wavelength of the resulting density perturbations in the gas is approximately equal to the thickness of the 1-mm dust layer.  We have been able to model the vortices for several hundred orbital periods, and find them to be stable.  The fractional oscillations in the gas radial density profile have an amplitude of $\sim 2\%$ (bottom panels of Fig.~\ref{Fig3}). On the contrary, the 50-cm dust grains (${\rm St} \approx 7$) are not coupled well enough to the gas to be entrained by the vortices and they quickly settle into the mid-plane. They correspond to the high-density regions located at the disc mid-plane in the bottom right panel of Fig.~\ref{Fig1}. Their radial migration, however, is stopped by the presence of the vortices. Local pressure maxima are created in the inter-vortex regions (bottom left panel of Fig.~\ref{Fig1}), and the 50-cm dust grains migrate into them and are subsequently trapped. 

In the lower panels of Fig.~\ref{Fig3}, the azimuthally-averaged radial density profiles of gas and dust at the mid-plane are presented. The presence of the vortices can be easily seen in the perturbed gas density profile, showing the aforementioned local minima and maxima. In the bottom left panel, the 1-mm dust and gas densities are compared, demonstrating the trapping of the small dust grains by the vortices (paradoxically corresponding to the pressure minima). This counterintuitive result can be understood as follows.  Due to the vertical component of gravity, the 1-mm dust grains tend to settle towards the disc mid-plane. However, they are also strongly dragged by the gas.  The 1-mm grains near the centres of the vortices are already near the mid-plane and the gas slowly circulates around the mid-plane, so there is nothing to move them away.  On the contrary, and as can be appreciated in the bottom panels of Fig.~\ref{Fig1}, gas in the inter-vortex regions strongly drags the 1-mm grains away from the mid-plane, lowering its mid-plane density (and producing the wave-like distribution of small dust). Conversely, in the bottom right panel of Fig. \ref{Fig3}, comparison of the 50-cm dust and gas radial density profiles shows that the large dust is concentrated in local pressure maxima. In the bottom left panel of Fig.~\ref{Fig2}, the azimuthally-averaged gas and 1-mm dust radial velocity profiles at the mid-plane are shown, illustrating the entrainment of the small dust by the gas vortices. In the bottom right panel, the azimuthally-averaged gas and 50-cm dust radial velocity profiles at the mid-plane are shown, demonstrating how the large dust grain migration is stopped at the local pressure maxima.  It should be noted that although large grains located between the pressure maxima still have substantial inward velocities, at this point in the calculation the vast majority of the large grains are already located at the pressure maxima (c.f.~Fig~\ref{Fig3}) and have negligible radial velocities.  The few large grains between pressure maxima will continue to migrate into the pressure maxima and be trapped.

The overall result is that the dust is agglomerated into rings throughout the disc, but that the radial locations of the rings are anti-correlated.  The small grains accumulate at the the centres of the toroidal vortices, while the large grains accumulate between the vortices at pressure maxima.  This clearly demonstrates that dust accumulation in protoplanetary discs, and its relation to the gas distribution, may be very complex. An obvious next step for future calculations is to investigate the behaviour of dust populations that have continuous grain size distributions (e.g. power-laws), but this is beyond the scope of this letter.

\section{Conclusions}
\label{conclusions}

We have investigated the evolution of a two-fold population of dust grains with sizes of 1 mm and 50 cm, within a gaseous protoplanetary disc. We find that the same instability as published by \cite{LB15a} is developed if the 1-mm dust grains are included in the simulation, regardless of the presence of 50-cm dust grains. After the development of the instability, dusty toroidal gaseous vortices are formed, entraining the 1-mm dust particles. The presence of such vortices creates a perturbed gas radial density profile, capable of stopping the migration of 50-cm dust grains by trapping them in pressure maxima created in the inter-vortex regions. The wavelength of the resulting mid-plane gas density perturbations is approximately equal to the thickness of the layer of small dust, with a relative amplitude of a few percent with respect to the unperturbed state. It is interesting to note the close resemblance of the present scenario with the model presented in \cite{Pea12} where a sinusoidal perturbation of the surface density profile of a disc was used to explain mm-observations of the outer parts of young protoplanetary discs and the inferred large grain sizes.  Although the physical prescription included in our simulations is still rather simple, the toroidal vortex instability may offer a simple and natural way to solve the dust migration problem.

\section*{Acknowledgments}
The figures were created using SPLASH \citep{Pri07}, a SPH visualization tool publicly available at http://users.monash.edu.au/$\sim$dprice/splash.  

This work was supported by the STFC consolidated grant ST/J001627/1, and by the European Research Council under the European Community's Seventh Framework Programme (FP7/2007-2013 grant agreement no. 339248). This work used the DiRAC Complexity system, operated by the University of Leicester IT Services, which forms part of the STFC DiRAC HPC Facility (www.dirac.ac.uk). This equipment is funded by BIS National E-Infrastructure capital grant ST/K000373/1 and  STFC DiRAC Operations grant ST/K0003259/1. DiRAC is part of the National E-Infrastructure. This work also used the University of Exeter Supercomputer, a DiRAC Facility jointly funded by STFC, the Large Facilities Capital Fund of BIS and the University of Exeter.

\bsp
\label{lastpage}
\end{document}